\newcommand*{\ATLASLATEXPATH}{./}
\newcommand\spacecite{\penalty700\ \cite}
\documentclass[preprint,texmf,texlive=2011,UKenglish,txfonts,siunitx=false]{atlasdoc}
\pdfoutput=1
\usepackage{atlasphysics}
\usepackage[biblatex=false,subfigure=true]{atlaspackage}
\usepackage{multirow}
\usepackage{cite}
\usepackage{bm}

\newcommand{\ptp}{\ensuremath{p_{\mathrm{T}}}}
\newcommand*\NumErrStat[2]{
    \ensuremath{#1 \pm #2\stat}
}
\newcommand*\NumErrStatSyst[3]{
    \ensuremath{#1 \pm #2\stat \pm #3\syst}
}

\graphicspath{{logos/}}

\AtlasTitle{Evidence of $\bm{W\gamma\gamma}$ production in $\bm{pp}$ collisions at $\bm{\sqrt{s}=8}$~TeV and limits on anomalous quartic gauge couplings with the ATLAS detector}

\PreprintIdNumber{CERN-PH-EP-2015-009}

\AtlasJournal{Physical Review Letters}

\AtlasAbstract{ %
This Letter reports evidence of triple gauge boson production
$pp\to W(\ell\nu)\gamma\gamma + X$, which is accessible for the first time with the 8~TeV LHC data set.
The fiducial cross section for this process is measured in a data sample
corresponding to an integrated luminosity of 20.3~fb$^{-1}$,
collected by the ATLAS detector in 2012.
Events are selected using the $W$ boson decay to $e\nu$
or $\mu\nu$ as well as requiring two isolated photons.
The measured cross section is used to set limits on anomalous quartic gauge couplings in the high diphoton mass region.
}

\begin{document}

In the Standard Model (SM), the self-couplings of the electroweak
gauge bosons are specified by the non-Abelian $SU(2) \times U(1)$ structure
of the electroweak sector. Since any deviation in the self-couplings from this
expectation indicates the presence of new physics phenomena at
unprobed energy scales, the measurement of the production of multiple
electroweak gauge bosons represents an important test of the SM.
This Letter presents a measurement of the triboson production
cross section, discussed in Ref.~\cite{bib:WggTheory},
where the W boson decays into $e\nu$ or $\mu\nu$~($W(\ell\nu)\gamma\gamma$),
and its sensitivity to anomalous quartic gauge couplings (AQGCs) $WW\gamma\gamma$. 
Such final states mainly come from events where the $W$ boson is produced in the 
hard interaction between the two partons, and the photons either originate from initial 
or final state radiation processes, or from triple or quartic gauge vertices together with the $W$ boson. 
The inclusive and exclusive cross sections are both measured.
The inclusive case has no restriction on the $W\gamma\gamma$ recoil system,
whereas the exclusive case includes a veto on events containing one or more jets.
Limits on AQGC parameters are set in the exclusive phase space with a diphoton mass larger than $300\, \GeV$.
Total and differential cross sections for the diboson production processes
$WW$, $WZ$, $ZZ$, $W\gamma$, and $Z\gamma$ have been reported previously by the
ATLAS~\cite{bib:ATLAS_WW, bib:ATLAS_WZ, bib:ATLAS_ZZ, bib:ATLAS_Wg}, CMS~\cite{bib:CMS_WW, bib:CMS_ZZ, bib:CMS_Wg},
D0~\cite{bib:D0_WZ_ZZ, bib:D0_Wg, bib:D0_aTGC}, and CDF~\cite{bib:CDF_WW, bib:CDF_VV, bib:CDF_Zg} collaborations,
including limits on anomalous triple gauge boson couplings.
Limits have been set on AQGCs by ATLAS~\cite{bib:ATLAS_ssWW}, CMS~\cite{bib:cmsqgc,bib:cmswvy}, the LEP experiments~\cite{bib:aleph,bib:delphi,bib:l3,bib:opal}, and D0~\cite{bib:D0_VV}.

ATLAS~\cite{ATLASdetector} is a multipurpose detector composed of
an inner tracking detector (ID) surrounded by a thin
superconducting solenoid providing a 2~T axial magnetic field,
electromagnetic (EM) and hadronic calorimeters, and a muon spectrometer (MS)
immersed in the magnetic field produced by a system of superconducting
toroids.
Events in this analysis are selected with triggers requiring the presence of one muon with a transverse momentum~($\ptp$) of more than $18\,\GeV$ and two electromagnetic objects with a transverse energy~($E_{\mathrm{T}}$) of more than $10\,\GeV$ each, with an efficiency of about $80\%$~\cite{bib:muoneff}, or three $E_{\mathrm{T}} > 15\,\GeV$ electromagnetic objects with an efficiency of more than $95\%$~\cite{bib:eleff}~\cite{etafootnote}. 
After applying data quality requirements, the data set corresponds to a total integrated
luminosity of $20.3 \pm 0.6~\rm{fb}^{-1}$~\cite{ATLASlumi2013}.

The main backgrounds to the $W(\ell\nu)\gamma\gamma$ process originate from processes with jets identified as
photons or leptons, referred to as fakes hereafter.
Data-driven techniques are used to estimate fakes, whereas Monte Carlo (MC)
simulation is used to estimate background sources with prompt leptons and photons and for the signal. 
The SHERPA~1.4.1 generator~\cite{bib:sherpa,bib:sherpa1,bib:sherpa2,bib:sherpa3} is used to model the signal with up to three partons in the final state.
SHERPA was also used to simulate the $Z\gamma$, $Z\gamma\gamma$, $WZ$, and $W(\tau\nu)\gamma\gamma$ backgrounds.
For the $Z\gamma$ background, the agreement between data and the MC prediction was assessed in $Z$-enriched control regions. 
The $t\bar{t}$, single top, and $WW$ processes are modeled by
MC@NLO~4.02~\cite{MCatNLO1,MCatNLO2}, interfaced to HERWIG~6.520~\cite{herwig}
for parton showering and fragmentation processes and to JIMMY~4.30~\cite{jimmy} for underlying event simulation.
The POWHEG~\cite{powheg} generator is used to simulate $ZZ$ production,
interfaced to PYTHIA~8.163~\cite{pythia} for parton showering and fragmentation.
The {\tt CT10} parton distribution function (PDF) set~\cite{Lai:2010vv} is used for all SHERPA, MC@NLO, and POWHEG samples.
The standard ATLAS detector simulation~\cite{atlassimu} based on GEANT4~\cite{geant4} is used. It includes multiple proton-proton
interactions per bunch crossing (pile-up) as observed in data.

The $W(\ell\nu)\gamma\gamma$ candidate events contain an isolated lepton
and missing transverse momentum ($E_{\mathrm{T}}^{\rm{miss}}$) from the undetected neutrino of the
leptonic $W$ decay, and two isolated photons (including also photons that have converted in electron-positron pairs within the ID volume). 
Muon candidates are identified, within pseudorapidity~\cite{etafootnote}
$|\eta|<2.4$, by associating complete tracks or track segments in the
MS with tracks in the ID~\cite{atlas_muons}.
Electron candidates are reconstructed within $|\eta| < 2.47$ as electromagnetic clusters associated to a track~\cite{atlas_electron},
whereas photons are reconstructed as electromagnetic clusters with $|\eta| < 2.37$~\cite{atlas_electron_calib}.
The calorimeter transition regions at $1.37 < |\eta| < 1.52$ are excluded for electrons and photons.
Identification criteria based on shower shapes in the EM calorimeter for photons, and additionally on tracking information
for electrons, referred to as ``tight'' in Refs.~\cite{ATLASPhotonIDConf, atlas_electron}, are used.
The $E_{\mathrm{T}}^{\rm{miss}}$ uses the  energy deposits in the calorimeters
within $|\eta|<4.9$ and the muons identified in the MS, as described in Ref.~\cite{METpaper}.
\begin{table*}[tb]
\begin{footnotesize}
\begin{center}
\resizebox{\columnwidth}{!} {
\begin{tabular}{c|cc|cc}
\hline
\hline
            & Electron channel & Muon channel & Electron channel & Muon channel \\
\hline
            & \multicolumn{2}{c}{$N_{\mathrm{jet}}\geq 0$} & \multicolumn{2}{c}{$N_{\mathrm{jet}} = 0$} \\
\hline
$W\gamma$j + $W$jj & \NumErrStatSyst{15.3}{4.8}{5.3}      & \NumErrStatSyst{30.5}{7.7}{6.8} & \NumErrStatSyst{5.8}{2.1}{2.0} & \NumErrStatSyst{14.4}{4.9}{4.9}  \\
$\gamma\gamma$ + jets    & \NumErrStatSyst{1.5}{0.6}{1.0} & \NumErrStatSyst{11.0}{4.0}{4.9} & \NumErrStatSyst{0.2}{0.2}{0.2} & \NumErrStatSyst{6.1}{3.5}{3.1}  \\
$Z\gamma$                 & \NumErrStat{11.2}{1.1}         & \NumErrStat{3.9}{0.2} & \NumErrStat{2.4}{0.5} & \NumErrStat{2.8}{0.2}  \\
Other backgrounds         & \NumErrStat{2.2}{0.6}         & \NumErrStat{6.7}{2.0} & \NumErrStat{0.3}{0.1} & \NumErrStat{1.1}{0.3}  \\
\hline
Total background  & \NumErrStatSyst{30.2}{5.0}{5.4}& \NumErrStatSyst{52.1}{8.9}{8.4} & \NumErrStatSyst{8.7}{2.2}{2.0} & \NumErrStatSyst{24.4}{6.0}{5.8}  \\
\hline
\hline
Data    &  $47$      &  $110$  & $15$ & $53$   \\
\hline
\hline
\end{tabular}
}
\end{center}
\end{footnotesize}
\caption{\label{tab:backgrounds}\small
The background composition in each channel is shown for the inclusive (left) and exclusive (right) cases. The $W\gamma$j + $W$jj and $\gamma\gamma$ + jets backgrounds are estimated using data-driven techniques, whereas the others are extracted from MC simulation. The number of candidate events in data passing the full selection is also shown.}
\end{table*}
Reconstructed muons, electrons, and photons are required to have $p_{\mathrm{T}}^{\mu,e,\gamma}>20\,\GeV$ and to
be isolated. Photons are considered isolated if the sum of calorimeter
transverse energy deposits in a cone of size $\Delta R = 0.4$ around the
candidate is smaller than $4\,\GeV$. The isolation is corrected for photon energy
leakage. The muon isolation is based on the sum of
the transverse momenta of ID tracks in a cone of size $\Delta R = 0.2$ which
must be below $0.15 \times \ptp^{\mu}$. For electrons, the calorimeter transverse
energy deposits and the sum of the transverse momenta of tracks in a cone of size
$\Delta R = 0.2$ must be below $0.2 \times \ptp^{e}$ and $0.15 \times \ptp^{e}$, respectively.
The lepton must also be compatible with originating from the primary vertex of the interaction,
which is taken to be the vertex with the largest $\Sigma p_{\mathrm{T}}^2$ of associated tracks.
$E_{\mathrm{T}}^{\rm{miss}}$ is required to exceed $25\,\GeV$.
The transverse mass of the $W$ boson~\cite{masstfootnote}
is required to be greater than $40\,\GeV$.
The two photons must be outside of their mutual isolation cones by requiring $\Delta R (\gamma, \gamma) > 0.4$.
To suppress the contribution from final-state radiation, the lepton and photons are required to have
$\Delta R (\ell, \gamma) > 0.7$.
Events containing a second reconstructed lepton are rejected to reduce background from
Drell-Yan events. In the electron channel, additional requirements are used to suppress
events in which one electron is misidentified as a photon (mainly originated from the Z$\gamma$ process):
the transverse momentum of the $e\gamma\gamma$ system is required to be greater than $30\,\GeV$,
and the invariant mass of the electron and the leading, subleading or both photons is required to be outside a $13$, $8$ or $15\, \GeV$
wide window around the $Z$ boson mass, respectively.
Exclusive events are defined with a veto on additional jets compared to the inclusive selection. Jets are reconstructed from clustered energy deposits in the
calorimeter using the anti-$k_t$ algorithm~\cite{antikt} with radius parameter $R=0.4$ and are required to have $\ptp>30\, \GeV$ and $|\eta|<4.4$.
Jets at $\Delta R < 0.3$ from the selected lepton
and photons are rejected.
In order to reduce pile-up effects,
for jets with $\ptp<50\, \GeV$ and $|\eta| < 2.4$, more than 50\% of the summed scalar $\ptp$ of tracks
within $\Delta R = 0.4$ of the jet axis must be from tracks associated to the primary vertex.

\begin{figure*}[htbp]
  \begin{center}
    \includegraphics[width=0.4\columnwidth]{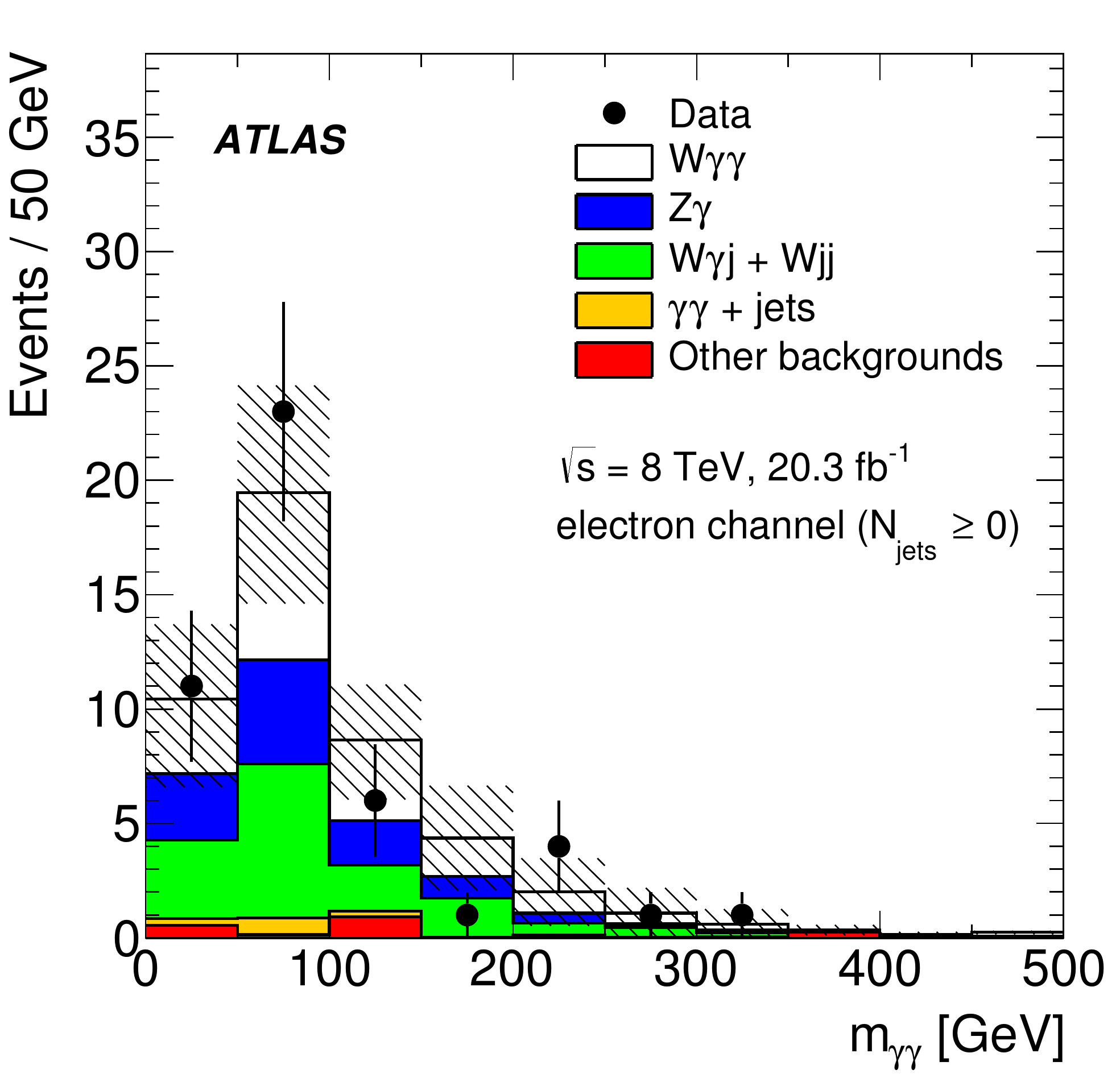}
    \includegraphics[width=0.4\columnwidth]{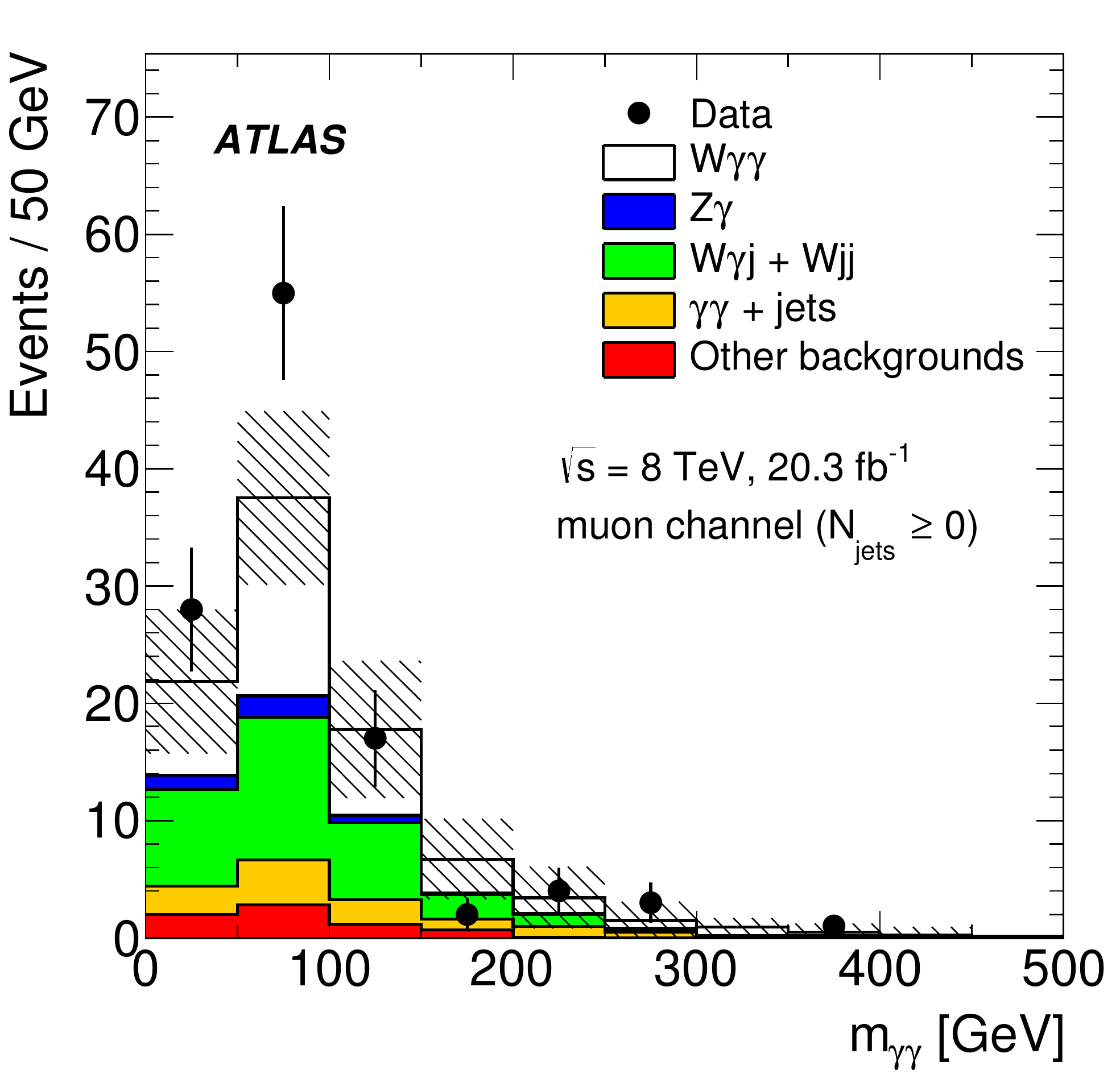}
    \caption{Diphoton invariant mass distribution in the electron (left) and muon (right) channels.
        The expected signal based on the SHERPA prediction is shown. The hashed areas show the total systematic and statistical uncertainty on the background estimate. }
    \label{fig:wgg_ggmass}
  \end{center}
\end{figure*}

Table~\ref{tab:backgrounds} shows the expected background as well as the
observation.
The background expectation alone is not sufficient to describe the data indicating the presence of signal events.
The fake-photon background from $W\gamma$j + $W$jj is estimated by performing
a two-dimensional template fit to the isolation energy distributions of the leading and subleading photons,
as described in Ref.~\cite{diphotonSM}.
Three background templates are obtained from data by reversing some of the
photon identification requirements based on shower shape; the signal templates are taken from MC simulation.
Contributions from events where a jet satisfies the electron identification criteria,
or the muon originates from heavy-flavor decays, i.e. from $\gamma\gamma$ + jets processes, are estimated
by using a two-dimensional sideband method constructed from the lepton isolation and $E_{\mathrm{T}}^{\rm{miss}}$ variables, as described
in Ref.~\cite{bib:ATLAS_Wg}.
The distribution of the diphoton invariant mass in the two channels is shown in Fig.~\ref{fig:wgg_ggmass}.
Alternative methods have been used to cross-check the estimate of the backgrounds coming from fakes, all of them provide consistent results. 
In the estimation of the fake-photon background,
systematic uncertainties arise from the limited number of events in the control regions,
the functional form used to describe the background isolation energy distribution,
the definition of the control region, the modeling of the signal in the
MC samples and the corresponding statistical uncertainty.
In the estimate of the fake-lepton background, systematic uncertainties related
to the control region definitions and the residual correlation of the discriminating variables are considered.

The fiducial cross sections $\sigma^{\mathrm{fid}}_{W\gamma\gamma}$ are obtained from a maximum-likelihood fit, similarly to
Ref.~\cite{bib:ATLAS_Wg}, for the electron channel, the muon channel, and the combination of the two
assuming lepton universality to determine the $W(\ell\nu)\gamma\gamma$ cross
section for a single lepton flavor.
They are measured in a phase space, defined in Table~\ref{tab:fiducialRegion}, close to that of the experimentally selected region.
Here $p_{\text{T}}^\nu$ is the transverse momentum of the neutrino and $\epsilon_{\rm{h}}^{\rm{p}}$ is
the fractional energy carried by the closest particle-level jet in a cone of $\Delta R =0.4$ around each photon direction.
\begin{table}[!ht]
  \centering
  \begin{tabular}{c}
    \hline
    \hline
    Definition of the fiducial region \\
    \hline
    $p_{\text{T}}^{\ell}>20\,\GeV$,
    $p_{\text{T}}^\nu>25\,\GeV$,
    $|\eta_{\ell}|<2.5$ \\
    $m_{\text{T}} > 40\,\GeV$\\
    $E_{\text{T}}^\gamma>20\,\GeV$,
    $|\eta^\gamma|<2.37$,
    iso. fraction $\epsilon_{\rm{h}}^{\rm{p}} < 0.5$\\
    $\Delta R(\ell,\gamma)>0.7$,
    $\Delta R(\gamma,\gamma)>0.4$,
    $\Delta R(\ell/\gamma, \mathrm{jet})>0.3$\\
    \\
    Exclusive: no anti-$k_t$ jets with $\ptp^{\mathrm{jet}}>30\,\GeV$, $|\eta^{\mathrm{jet}}|<4.4$\\
    \hline
    \hline
  \end{tabular}
  \caption{Definition of the fiducial region for which the cross section is
           evaluated.}
  \label{tab:fiducialRegion}
\end{table}

The efficiency of the signal selection and the small acceptance correction due to the extrapolation over the calorimeter transition region and
to $|\eta| = 2.5$ for the leptons are taken into account in the procedure. The acceptance correction factors are 0.83 and 0.90 in the electron and muon channel, respectively.
The combined efficiency and acceptance correction
amounts to $(19.6 \pm 0.5)$\%
and $(40.4 \pm 0.7)$\% in the electron and muon channels in the inclusive case, and to $(15.1 \pm 0.7)$\% and $(39.7 \pm 1.0)$\% in the exclusive case.
The given uncertainties are statistical only.
Corrections are applied to account for small differences between data and MC simulation in lepton, photon, and jet efficiencies, momentum scale and resolution, additional $pp$ interactions, and beam-spot position.

Systematic uncertainties on the cross section are accounted for by introducing nuisance
parameters in the likelihood which modify the signal and background expected yields.
Correlations between systematic uncertainties in the two channels are accounted for in the combined fit.
When combining the two channels, the dominant systematic uncertainties in the inclusive and exclusive cross-section measurements are 14\% and 23\% from the data-driven background estimates, 5\% to 7\% from the jet energy scale, and 3\% from the luminosity.
Other systematic uncertainties considered stem from the electromagnetic and muonic energy scale and resolution, the object reconstruction, the pileup description, and the trigger efficiency. These are found to have a minor impact, below 3\%.
Theoretical uncertainties on the signal modeling, affecting only the acceptance extrapolation, are negligible.
The measured cross sections are shown in Table~\ref{table:xsecRes}.
The significance after combining the two channels is larger than $3\sigma$ in the inclusive case.
The measurements in the electron and muon channels are compatible within $1\sigma$.
\begin{table*}[tb]
\centering
 \begin{tabular}{c|c|c}
       \hline
       \hline
       & $\sigma^{\rm{fid}}$ [fb] & $\sigma^{\rm{MCFM}}$ [fb]\\
       \hline
       Inclusive ($N_{\mathrm{jet}} \geq 0$) \\
       \hline
       \rule{0pt}{2.5ex}
       $\mu\nu\gamma\gamma$  & $7.1~~^{+1.3}_{-1.2}$ (stat.) $\pm 1.5$ (syst.) $\pm$0.2 (lumi.) & \multirow{3}{*}{$2.90 \pm 0.16$} \\
       $e\nu\gamma\gamma$    & $4.3~~^{+1.8}_{-1.6}$ (stat.) $^{+1.9}_{-1.8}$ (syst.) $\pm$0.2 (lumi.) &  \\
       $\ell\nu\gamma\gamma$ & $6.1~~^{+1.1}_{-1.0}$ (stat.) $\pm 1.2$ (syst.) $\pm$0.2 (lumi.) & \\
       \hline
       Exclusive ($N_{\mathrm{jet}} = 0$) \\
       \hline
       \rule{0pt}{2.5ex}
       $\mu\nu\gamma\gamma$  & $3.5\pm 0.9$ (stat.) $^{+1.1}_{-1.0}$ (syst.) $\pm$0.1 (lumi.)  & \multirow{3}{*}{$1.88 \pm 0.20 $}\\
       $e\nu\gamma\gamma$    & $1.9~~^{+1.4}_{-1.1}$ (stat.) $^{+1.1}_{-1.2}$ (syst.) $\pm$0.1 (lumi.)         & \\
       $\ell\nu\gamma\gamma$ & $2.9~~^{+0.8}_{-0.7}$ (stat.) $^{+1.0}_{-0.9}$ (syst.) $\pm$0.1 (lumi.)  & \\
       \hline
       \hline
\end{tabular}
\caption{Measurement of the $pp\rightarrow \ell\nu\gamma\gamma + X$ inclusive and exclusive fiducial cross sections. }
  \label{table:xsecRes}
\end{table*}

The SM prediction for the $W(\ell\nu)\gamma\gamma$ cross section is calculated with the parton-level MC program MCFM~\cite{bib:mcfm} at next-to-leading order (NLO).
The calculations are performed using the MCFM default electroweak parameters~\cite{bib:mcfmmanual}
and the {\tt CT10} PDF set.
The renormalization and factorization scales are set to the invariant mass of the
$\ell\nu\gamma\gamma$ system. The fragmentation of quarks and gluons to photons
is included using the fragmentation function {\tt GdRG$\_$LO}\cite{bib:GdRG}.
The kinematic requirements at parton level match the fiducial acceptance of Table~\ref{tab:fiducialRegion}.

In addition to the inclusive prediction, an exclusive cross section is obtained by vetoing events with an additional jet emission. To account for the difference between jets defined at parton and particle levels, a correction factor of about 0.87 in the exclusive case is computed and applied to the prediction as documented in Ref.~\cite{bib:ATLAS_Wg}.
Uncertainties on the two predictions include the effect of varying independently the renormalization and factorization scales by factors of $0.5$ and $2.0$, evaluating the {\tt CT10} PDF error sets scaled to the 68\% confidence level (CL), the uncertainties on quark or gluon fragmentation to a photon, and the parton
to particle correction factors.
The predictions for $W(\ell\nu)\gamma\gamma$ production are compared to the measured cross sections in Table~\ref{table:xsecRes}.
The measured cross section is higher by $1.9\sigma$ in the inclusive case, while better agreement is seen in the exclusive case,
similar to the measurement of $W\gamma$ and $Z\gamma$ in Ref.~\cite{bib:ATLAS_Wg}. In the case of $Z\gamma$ and $W\gamma$, 
higher order corrections were calculated to be smaller for the exclusive compared to the inclusive case~\cite{bib:Grazzini}. 
As the process $W\gamma\gamma$ has similar properties, the exclusive measurement is expected to be in better 
agreement with the theoretical prediction than the inclusive one. Therefore, in the following, the exclusive measurement 
will be used for the AQGC limits setting, as done in Ref.~\cite{bib:ATLAS_Wg}.

The AQGCs are introduced as dimension-8 operators following the formalism
defined in the Appendix of Ref.~\cite{bib:eboli}. While many
operators give rise to anomalous couplings of the form $WW\gamma\gamma$,
this study is restricted to $f_{\mathrm{T0}}/\Lambda^4$, $f_{\mathrm{M2}}/\Lambda^4$, and $f_{\mathrm{M3}}/\Lambda^4$, where $\Lambda$ represents the scale at which new physics appears, and $f$ the coupling of the respective operator.
The $W\gamma\gamma$ final state is expected to be particularly sensitive to the $\mathrm{T0}$ operator,
whereas the other two operators can be related to the parameters of the dimension-6 operators used at LEP~\cite{bib:aleph,bib:delphi,bib:l3,bib:opal} and by CMS~\cite{bib:cmsqgc} via the transformations described in Ref.~\cite{bib:AQGCconversion}.
To preserve unitarity up to high energy scales, a form factor is introduced 
which depends on the energy, the form factor scale $\Lambda_{\mathrm{FF}}$ and an exponent $n$, following the formalism described in Refs.~\cite{bib:AQGCmethod1, bib:AQGCmethod2}. The scale $\Lambda_{\mathrm{FF}}$ is independent of the new physics scale $\Lambda$~\cite{bib:FFscale}.
The largest form factor scale ensuring unitarity for this
process at $\sqrt{s} = 8\,\TeV$, calculated using the VBFNLO generator~\cite{bib:calc_formfac,bib:VBFNLO1,bib:VBFNLO2,bib:VBFNLO3},
is given by $n=2$ and $\Lambda_{\rm{FF}} = 600\,\GeV$ for $f_{\mathrm{T0}}/\Lambda^4$,
and $\Lambda_{\rm{FF}} = 500\,\GeV$ for $f_{\mathrm{M2}}/\Lambda^4$ and $f_{\mathrm{M3}}/\Lambda^4$.

Deviations from the SM prediction for the AQGC parameters, which are predicted to be zero, lead to an excess of events with high diphoton invariant mass.
The phase space to study AQGCs was optimized using the expected significance calculated on simulated events. The optimal phase space was found to be the exclusive selection with the additional requirement of $m_{\gamma\gamma}>300\,\GeV$. 
The SM backgrounds in this region are determined from a fit to the observed $m_{\gamma\gamma}$ distribution.
The expected SM background is \NumErrStatSyst{0.01}{0.03}{0.20} (\NumErrStatSyst{0.02}{0.05}{0.46}) events in the electron (muon) channel, where uncertainties include systematic effects due to the extrapolation procedure, depending on the modelling of the spectrum and on the initial background estimate used in the extrapolation. No events are observed in the high-mass region.

\begin{table}[htb]
  \centering
  \begin{tabular}{c||c|cc}
    \hline
    \hline
    & & Observed [$\TeV^{-4}$] & Expected [$\TeV^{-4}$]\\
    \hline
    \rule{0pt}{2.5ex}
    \multirow{3}{*}{$n=0$} & $f_{\mathrm{T0}}/\Lambda^4$ & [$-0.9$, 0.9] $\times$ 10$^{2}$ &  [$-1.2$, 1.2] $\times$ 10$^{2}$\\
      & $f_{\mathrm{M2}}/\Lambda^4$& [$-0.8$, 0.8] $\times$ 10$^{4}$ & [$-1.1$, 1.1] $\times$ 10$^{4}$\\
      & $f_{\mathrm{M3}}/\Lambda^4$ & [$-1.5$, 1.4] $\times$ 10$^{4}$ & [$-1.9$, 1.8] $\times$ 10$^{4}$\\
      \hline
      \rule{0pt}{2.5ex}
      \multirow{3}{*}{$n=1$} & $f_{\mathrm{T0}}/\Lambda^4$ & [$-7.6$, 7.3] $\times$ 10$^{2}$ &  [$-9.6$, 9.5] $\times$ 10$^{2}$\\
			  & $f_{\mathrm{M2}}/\Lambda^4$ & [$-4.4$, 4.6] $\times$ 10$^{4}$ &  [$-5.7$, 5.9] $\times$ 10$^{4}$\\
        & $f_{\mathrm{M3}}/\Lambda^4$ & [$-8.9$, 8.0] $\times$ 10$^{4}$ & [$-11.0$, 10.0] $\times$ 10$^{4}$\\
        \hline
        \rule{0pt}{2.5ex}
        \multirow{3}{*}{$n=2$} & $f_{\mathrm{T0}}/\Lambda^4$ & [$-2.7$, 2.6] $\times$ 10$^{3}$ & [$-3.5$, 3.4] $\times$ 10$^{3}$\\
				  & $f_{\mathrm{M2}}/\Lambda^4$ & [$-1.3$, 1.3] $\times$ 10$^{5}$ & [$-1.6$, 1.7] $\times$ 10$^{5}$\\
          & $f_{\mathrm{M3}}/\Lambda^4$ & [$-2.9$, 2.5] $\times$ 10$^{5}$ & [$-3.7$, 3.3] $\times$ 10$^{5}$\\
    \hline
    \hline
  \end{tabular}
  \caption{Observed and expected 95\% CL limits obtained
    for the $f_{\mathrm{T0}}/\Lambda^4$, $f_{\mathrm{M2}}/\Lambda^4$ and $f_{\mathrm{M3}}/\Lambda^4$ AQGC parameters for the combination of the two channels. The values of $n = 0,\,1,\,2$ are the exponential choices of the form factor, $\Lambda_{\rm{FF}}$ is fixed to $600\,\GeV$ for $f_{\mathrm{T0}}/\Lambda^4$ and to $500\,\GeV$ for the other parameters. The $n=0$ choice produces the limits without the form factor applied.}
  \label{tab:AQGCresults}
\end{table}
The cross-section prediction as a quadratic function of the AQGC parameters is obtained by using VBFNLO~\spacecite{bib:calc_formfac,bib:VBFNLO1,bib:VBFNLO2,bib:VBFNLO3}.
For SM couplings VBFNLO agrees with MCFM. The limits on the AQGC parameters are extracted with a frequentist profile likelihood test~\cite{CLpaper}, using the methodology of Ref.~\cite{bib:ATLAS_Wg}. The expected and observed limits at 95\% CL on the AQGC parameters are shown in Table~\ref{tab:AQGCresults}
for different values of $n$. The limits on $f_{\mathrm{M2}}/\Lambda^4$ and $f_{\mathrm{{M3}}}/\Lambda^4$ improve on the previous results from LEP~\cite{bib:aleph,bib:delphi,bib:l3,bib:opal} and D0~\cite{bib:D0_VV}, but are less stringent than those from CMS~\cite{bib:cmsqgc,bib:cmswvy}. The limit on $f_{\mathrm{T0}}/\Lambda^4$ is tighter than the previous limit published by CMS~\cite{bib:cmswvy,AQGCparamfootnote}. This can be explained by the fact that $f_{\mathrm{T0}}/\Lambda^4$ is especially sensitive to transversely polarized $W$
bosons, which are favored in the present study~\cite{bib:eboli}.

In summary, evidence for the $W(\ell\nu)\gamma\gamma$ process is reported for the first time.
The significance of the inclusive production cross section is larger than $3\, \sigma$.
The measured cross sections are in agreement within uncertainties with NLO SM predictions calculated
with MCFM.
Limits are set at 95\% CL on the AQGC parameters,
in particular improving the limit on $f_{\mathrm{T0}}/\Lambda^4$.

\vskip 5mm


We thank CERN for the very successful operation of the LHC, as well as the
support staff from our institutions without whom ATLAS could not be
operated efficiently.

We acknowledge the support of ANPCyT, Argentina; YerPhI, Armenia; ARC,
Australia; BMWFW and FWF, Austria; ANAS, Azerbaijan; SSTC, Belarus; CNPq and FAPESP,
Brazil; NSERC, NRC and CFI, Canada; CERN; CONICYT, Chile; CAS, MOST and NSFC,
China; COLCIENCIAS, Colombia; MSMT CR, MPO CR and VSC CR, Czech Republic;
DNRF, DNSRC and Lundbeck Foundation, Denmark; EPLANET, ERC and NSRF, European Union;
IN2P3-CNRS, CEA-DSM/IRFU, France; GNSF, Georgia; BMBF, DFG, HGF, MPG and AvH
Foundation, Germany; GSRT and NSRF, Greece; RGC, Hong Kong SAR, China; ISF, MINERVA, GIF, I-CORE and Benoziyo Center, Israel; INFN, Italy; MEXT and JSPS, Japan; CNRST, Morocco; FOM and NWO, Netherlands; BRF and RCN, Norway; MNiSW and NCN, Poland; GRICES and FCT, Portugal; MNE/IFA, Romania; MES of Russia and NRC KI, Russian Federation; JINR; MSTD,
Serbia; MSSR, Slovakia; ARRS and MIZ\v{S}, Slovenia; DST/NRF, South Africa;
MINECO, Spain; SRC and Wallenberg Foundation, Sweden; SER, SNSF and Cantons of
Bern and Geneva, Switzerland; NSC, Taiwan; TAEK, Turkey; STFC, the Royal
Society and Leverhulme Trust, United Kingdom; DOE and NSF, United States of
America.

The crucial computing support from all WLCG partners is acknowledged
gratefully, in particular from CERN and the ATLAS Tier-1 facilities at
TRIUMF (Canada), NDGF (Denmark, Norway, Sweden), CC-IN2P3 (France),
KIT/GridKA (Germany), INFN-CNAF (Italy), NL-T1 (Netherlands), PIC (Spain),
ASGC (Taiwan), RAL (UK) and BNL (USA) and in the Tier-2 facilities
worldwide.

\clearpage

\begin{flushleft}
{\Large The ATLAS Collaboration}

\bigskip

G.~Aad$^{\rm 85}$,
B.~Abbott$^{\rm 113}$,
J.~Abdallah$^{\rm 152}$,
S.~Abdel~Khalek$^{\rm 117}$,
O.~Abdinov$^{\rm 11}$,
R.~Aben$^{\rm 107}$,
B.~Abi$^{\rm 114}$,
M.~Abolins$^{\rm 90}$,
O.S.~AbouZeid$^{\rm 159}$,
H.~Abramowicz$^{\rm 154}$
% ...
\bigskip
\\
% ...
$^{11}$ Institute of Physics, Azerbaijan Academy of Sciences, Baku, Azerbaijan\\
% ...
$^{85}$ CPPM, Aix-Marseille Universit{\'e} and CNRS/IN2P3, Marseille, France\\
% ...
$^{90}$ Department of Physics and Astronomy, Michigan State University, East Lansing MI, United States of America\\
% ...
$^{107}$ Nikhef National Institute for Subatomic Physics and University of Amsterdam, Amsterdam, Netherlands\\
% ...
$^{113}$ Homer L. Dodge Department of Physics and Astronomy, University of Oklahoma, Norman OK, United States of America\\
$^{114}$ Department of Physics, Oklahoma State University, Stillwater OK, United States of America\\
% ...
$^{117}$ LAL, Universit{\'e} Paris-Sud and CNRS/IN2P3, Orsay, France\\
% ...
$^{152}$ Institute of Physics, Academia Sinica, Taipei, Taiwan\\
% ...
$^{154}$ Raymond and Beverly Sackler School of Physics and Astronomy, Tel Aviv University, Tel Aviv, Israel\\
% ...
$^{159}$ Department of Physics, University of Toronto, Toronto ON, Canada\\
% ...
\end{flushleft}


\begin{thebibliography}{99}
\bibitem{etafootnote}
  {ATLAS uses a right-handed coordinate system with its origin at the nominal interaction point (IP) in the center of the detector and the $z$-axis along the beam pipe. The $x$-axis points from the IP to the center of the LHC ring, and the $y$-axis points upward. Cylindrical coordinates $(r,\phi)$ are used in the transverse $(x,y)$ plane, $\phi$ being the azimuthal angle around the beam pipe. The pseudorapidity is defined in terms of the polar angle $\theta$ as $\eta=-\ln\tan(\theta/2)$. The distance $\Delta R$ in the $(\eta,\phi)$ space is defined as $\Delta R = \sqrt{({\Delta \eta})^2 + ({\Delta \phi})^2}$.}
\bibitem{masstfootnote}
  {The transverse mass of the $W$ is defined, using the lepton ($\ell$) and neutrino ($\nu$) $p_{\mathrm{T}}$ and $\phi$, as $m_{\rm{T}} = \sqrt{2 \ptp^{\ell} \ptp^{\nu} (1-\rm{cos}(\phi_{\ell}-\phi_{\nu}))}$.}
\bibitem{AQGCparamfootnote}
  {VBFNLO uses a slightly different definition of the field strength tensors than Ref.~\cite{bib:eboli}. Therefore, the couplings $f$ need to be scaled for comparison. In the case of $f_{\mathrm{T0}}$, the scale factor is $ g^{-4}$, where $g$ is the $SU(2)$ gauge coupling~\cite{bib:calc_formfac}.}
\end{thebibliography}
\end{document}